\documentclass[12pt]{article}
\usepackage[latin1]{inputenc}

\usepackage{amsmath}
\usepackage{amsfonts}
\usepackage{amssymb}
\usepackage{graphicx}
\usepackage{geometry}
\usepackage{amssymb,epsfig,subfigure}
\usepackage{hyperref}

\def\simge{
    \mathrel{\rlap{\raise 0.511ex 
        \hbox{$>$}}{\lower 0.511ex \hbox{$\sim$}}}}

\def\simle{
    \mathrel{\rlap{\raise 0.511ex 
        \hbox{$<$}}{\lower 0.511ex \hbox{$\sim$}}}}

\makeatletter
\renewcommand\section{\@startsection {section}{1}{\z@}%
                                 {-3.5ex \@plus -1ex \@minus -.2ex}
                                   {2.3ex \@plus.2ex}%
                                   {\normalfont\large\bfseries}}
\renewcommand\subsection{\@startsection{subsection}{2}{\z@}%
                                   {-3.25ex\@plus -1ex \@minus -.2ex}%
                                     {1.5ex \@plus .2ex}%
                                     {\normalfont\bfseries}}
\renewcommand\subsubsection{\@startsection{subsubsection}{3}{\z@}%
                                   {-3.25ex\@plus -1ex \@minus -.2ex}%
                                     {1.5ex \@plus .2ex}%
                                     {\normalfont\itshape}}
\makeatother

\def\pplogo{\vbox{\kern-\headheight\kern -29pt
\halign{##&##\hfil\cr&{\ppnumber}\cr\rule{0pt}{2.5ex}&\ppdate\cr}}}
\makeatletter
\def\ps@firstpage{\ps@empty \def\@oddhead{\hss\pplogo}%
  \let\@evenhead\@oddhead 
}
\def\maketitle{\par
 \begingroup
 \def\thefootnote{\fnsymbol{footnote}}
 \def\@makefnmark{\hbox{$^{\@thefnmark}$\hss}}
 \if@twocolumn
 \twocolumn[\@maketitle]
 \else \newpage
 \global\@topnum\z@ \@maketitle \fi\thispagestyle{firstpage}\@thanks
 \endgroup
 \setcounter{footnote}{0}
 \let\maketitle\relax
 \let\@maketitle\relax
 \gdef\@thanks{}\gdef\@author{}\gdef\@title{}\let\thanks\relax}
\makeatother

\numberwithin{equation}{section}

\renewcommand{\dag}{\dagger}

\newcommand{\be}{\begin{equation}}
\newcommand{\bea}{\begin{eqnarray}}
\newcommand{\ee}{\end{equation}}
\newcommand{\eea}{\end{eqnarray}}
\newcommand\beq{\begin{equation}}
\newcommand\eeq{\end{equation}}

\newcommand{\mc}{\mathcal}

\newcommand{\tr}{{\rm tr}}

\begin{document}

\setcounter{page}0
\def\ppnumber{\vbox{\baselineskip14pt
}}
\def\ppdate{\footnotesize{SLAC-PUB-15201, SU/ITP-12/21}} \date{}

\author{Xi Dong, Bart Horn,  Eva Silverstein, Gonzalo Torroba\\
[7mm]
{\normalsize \it Stanford Institute for Theoretical Physics }\\
{\normalsize  \it Department of Physics, Stanford University }\\
{\normalsize \it Stanford, CA 94305, USA}\\
[3mm]
{\normalsize \it Theory Group, SLAC National Accelerator Laboratory}\\
{\normalsize  \it Menlo Park, CA 94025, USA}\\
[3mm]}

\bigskip
\title{\bf  Perturbative Critical Behavior from Spacetime Dependent Couplings
\vskip 0.5cm}
\maketitle

\begin{abstract}

We find novel perturbative fixed points by introducing mildly spacetime-dependent couplings into otherwise marginal terms. In four-dimensional QFT, these are physical analogues of the small-$\epsilon$ Wilson-Fisher fixed point.  Rather than considering $4-\epsilon$ dimensions, we stay in four dimensions but introduce couplings whose leading spacetime dependence is of the form $\lambda x^\kappa \mu^\kappa$, with a small parameter $\kappa$ playing a role analogous to $\epsilon$.  We show, in $\phi^4$ theory and in QED and QCD with massless flavors, that this leads to a critical theory under perturbative control over an exponentially wide window of spacetime positions $x$.  The exact fixed point coupling $\lambda_*(x)$ in our theory is identical to the running coupling of the translationally invariant theory, with the scale replaced by $1/x$.  Similar statements hold for three-dimensional $\phi^6$ theories and two-dimensional sigma models with curved target spaces.  We also describe strongly coupled examples using conformal perturbation theory.  
         
\end{abstract}
\bigskip
\newpage

\tableofcontents

\vskip 1cm

\section{Introduction}\label{sec:intro}

Quantum field theory provides a framework describing a rich set of phases.  
Scale invariant theories and their relevant perturbations play a central role, describing phase transitions and the behavior of physical quantities in the approach to criticality.  Basic quantum corrections in field theory manifest themselves beautifully in measured critical exponents.  RG fixed points are also important in formal studies of quantum field theory.     

Some classes of fixed points are amenable to perturbation theory, such as three dimensional gauge theories with large numbers of flavors \cite{Appelquist}, the three dimensional $O(N)$ model at large-$N$ \cite{ON}, and four dimensional Banks-Zaks fixed points \cite{BanksZaks}.  A prominent example formally is the Wilson-Fisher fixed point in $4-\epsilon$ dimensions~\cite{WF}.  This is perturbative at small $\epsilon$, with a classical negative $\beta$ function of order $\epsilon$ cancelling against the leading quantum correction.   Small $\epsilon$ is unphysical, but it led to progress on the three dimensional case with $\epsilon\to 1$:\ expanding in the loop factor and resumming leads to reasonable agreement with data on spin systems for several different universality classes.

In this paper, we present a new class of perturbatively controlled scale invariant theories.
These arise in physical (integer) dimensionalities $n$ and do not require a large flavor or color expansion.   Rather than considering $n-\epsilon$ dimensions, we stay in $n$ dimensions but start with couplings which at leading order take the form $\lambda x^\kappa \mu^\kappa$.  The small parameter $\kappa$ plays a role analogous to $\epsilon$:\  it produces a small classical contribution to the $\beta$ function for otherwise marginal couplings.  This classical contribution plays off against the quantum corrections (which would otherwise lead to marginal (ir)relevance of the coupling) to cancel the $\beta$ function.  For an exponentially wide range of positions $x$, $e^{-1/\kappa} \ll x\mu \ll e^{1/\kappa}$, we only require tuning one real parameter in $\lambda(x)$ to obtain a scale invariant theory to very good approximation.  
We show more generally that the exact fixed point coupling $\lambda_*(x)$ takes the same functional form as the running coupling of the unperturbed (translationally invariant) theory,  with $1/x$ playing the role of the scale in the running coupling.  In simple field theories, this is perturbatively controlled over an exponentially large range of spacetime positions $x$.     
It is worth emphasizing that our $\beta$ functions are defined with respect to a rescaling of all coordinates (as opposed to just coordinate differences), as in \cite{OsbAlbion}.  
 
Our examples include purely perturbative quantum field theories with minimal field content, including $\phi^4$ theory and massless QED and QCD in four dimensions, $\phi^6$ theory in three dimensions, and sigma models with curved target spaces in two dimensions.  Additional classes come from conformal perturbation theory away from strongly coupled CFTs, including examples controlled by supersymmetry or AdS/CFT.  Our original motivation came from ~\cite{Dong:2012ena}\, where we found that time dependent couplings in large $N$ CFTs can strongly modify long-distance physics and unitarity bounds, features which are important in holographic duals of time dependent gravitational systems \cite{FRW}.  It will be interesting to consider potential applications of this structure in physical systems, particularly since the effect arises in relatively simple non-supersymmetric theories without large numbers of species.

It is worth noting that our theories exhibit scale without conformal invariance, as we do not require translation invariance as in \cite{scaleconf}.  Another setting in which spacetime dependence arises is in the theory of disorder; see \cite{disorder}\ and references therein for a recent RG treatment.               

\vskip 2mm

This paper is organized as follows.    
We begin in \S \ref{sec:4dscalar} with the simplest example, corresponding to the $\lambda \phi^4$ theory in four dimensions with a spacetime dependent coupling $\lambda(x)\approx \lambda x^\kappa \mu^\kappa$. This theory is weakly coupled, which will allow us to illustrate our idea in a completely explicit way. The result is a physical analogue of the Wilson-Fisher fixed point, where the role of the small $\epsilon$ parameter is played by $\kappa$. After describing this in a way that captures the leading effect, we analyze the complete $x$-dependence of the fixed point coupling using the renormalization group equations, arriving at a simple result in terms of the running coupling of the translation-invariant theory.  In \S \ref{sec:kappaQED} we find similar fixed points in four-dimensional gauge theories with massless flavors, which arise from adding spacetime dependence to the gauge coupling, and discuss the behavior of the potential between charges as spacetime coordinates are rescaled.
We discuss three dimensional examples, relevant to possible condensed matter applications, in \S \ref{sec:3d}. Moving down to two dimensions, in \S \ref{sec:sigma-model} we add spacetime dependence to sigma models on curved spaces, finding that this gives rise to fixed points which describe Einstein manifolds in the target space.    
We next introduce a more general framework for our mechanism, and analyze spacetime-dependent deformations of supersymmetric theories in \S \ref{sec:susy}.  This softens the mass renormalization and also enables us to construct some illustrative strongly coupled versions of our mechanism. We end in \S \ref{sec:discussion}\ with brief remarks on possible applications. 

\section{$4d$ scalar fields and perturbative fixed points}\label{sec:4dscalar}

To begin, let us consider scalar field theory with a classical action
\beq\label{kappafour}
S=\int d^d x \left\{\frac{1}{2} (\partial\phi)^2 - \frac{\lambda x^\kappa \mu^{\kappa+\epsilon}}{4!} \phi^4 \right\}
\eeq
where $\epsilon=4-d$ serves as a regulator (to be taken to zero in the case of $4d$ QFT).  
Here $x$ could be in a spacelike or a timelike direction.\footnote{The spatial case is simpler for our present purposes as it avoids non-adiabatic effects, but both are interesting.}  Let us take $\kappa$ to be a small positive parameter, and start by working over a wide but limited range of spacetime scales satisfying
\beq\label{xbound}
e^{-1/\kappa}  \ll x\mu \ll  e^{1/\kappa}.
\eeq
In this section, we will first determine the leading effect of the $x$-dependence introduced by $\kappa\simge 0$ on the behavior of this theory under changes of scale.  Then we will complete our analysis by extending it to a general spacetime dependent coupling $\lambda(x)$.    

We can see from (\ref{kappafour}) that $\kappa$ plays a role which is somewhat similar to $\epsilon$.  However, an important difference is that $\epsilon$ functions as a UV cutoff and $\kappa$ does not.  This means that for the theory (\ref{kappafour}), we will also restrict to relative distances $\Delta x$ longer than the scale of the Landau pole:
\beq\label{Landau}
\Delta x\mu \gg e^{-c/\lambda}
\eeq
\noindent where $c$ is some positive constant of order one.  Similar comments will apply below to other examples based on marginally irrelevant couplings.  

The coupling $\lambda$ classically runs as 
\beq\label{classflow}
\lambda' = \lambda b^{-\kappa-\epsilon},
\eeq
under a rescaling of coordinates $x^\mu\to x^\mu/b$, growing stronger in the infrared.  
At one loop, the standard spacetime-independent $\lambda\phi^4$ theory develops a logarithmic contribution pushing $\lambda$ toward weaker coupling in the infrared, and this occurs in our theory as well for $\lambda\gg\kappa$.  These two effects will play off each other to produce a fixed point at a coupling $\lambda_*$ of order $\kappa$ to very good approximation over the range of scales (\ref{xbound}), and we will also solve for the complete $x$-dependence required to obtain an exact fixed point.           
Before analyzing that, let us briefly review the standard case introduced by Wilson and Fisher \cite{WF}.

\subsection{Wilson-Fisher review}\label{subsec:WF}

For $\kappa=0$, at one-loop order there appears the standard logarithmic contribution to the running of the $\phi^4$ coupling, given by $\frac{3\lambda^2}{16\pi^2} \,\log(b)$, and a mass correction which we will tune away.      
Combining this with (\ref{classflow}), the theory with $0<\epsilon\ll 1$ and $\kappa=0$ exhibits a perturbative fixed point \cite{WF}.  
One can find the fixed point coupling and establish its validity to all orders using various methods, such as those explained pedagogically in e.g.\ \cite{QFT,Kleinert}.  

In the (modified) minimal subtraction renormalization scheme, the bare coupling $\tilde\lambda_0$ appearing in the Lagrangian as $-\int d^d x\tilde\lambda_0\phi^4/4!$ takes the form
\beq\label{minsub}
\tilde\lambda_0 =\mu^\epsilon \left[\lambda+\frac{3\lambda^2}{16\pi^2}\frac{1}{\epsilon}+\sum_{n=3}^\infty\lambda^n \sum_{i=1}^{n-1} \frac{c_{ni}}{\epsilon^i}\right]
\eeq       
with the $c_{ni}$'s determined by requiring that the poles in $\epsilon$ cancel order by order in $\lambda$.  Noting that $\tilde\lambda_0$ depends only on $\epsilon$ and not on $\mu$, one has 
\beq\label{bare}
0=\mu\frac{\partial\tilde\lambda_0}{\partial\mu}= \epsilon \mu^\epsilon \left[\dots\right] + \mu^\epsilon \left(\mu\frac{\partial\lambda}{\partial\mu}\right)\partial_\lambda\left[\dots\right], 
\eeq   
where $[\dots]$ is the bracketed quantity in (\ref{minsub}).  
This allows us to solve for the $\beta$ function 
\beq\label{betaWF}
\beta=\mu\frac{\partial\lambda}{\partial\mu}=-\epsilon\lambda + \beta_{4d}(\lambda).
\eeq
Here $\beta_{4d}(\lambda)=3\lambda^2/16\pi^2 + \dots$ takes the form of a perturbative expansion in powers of $\lambda$ with no dependence on $\epsilon$, the coefficients $c_{ni}$ having been chosen to cancel all of the poles.  Aside from the classical term, the surviving infinitely many terms in the $\beta$ function come from products of a $1/\epsilon$ pole and the term proportional to $\epsilon$ appearing on the RHS of (\ref{bare}).    
From this one can immediately see that there is a fixed point at $\lambda_{*WF}=16\pi^2\epsilon/3+{\cal O}(\epsilon^2)$. 

This fixed point is under perturbative control for $\epsilon\ll 1$, but being away from an integer dimension it is not amenable to physical applications.  Pushing $\epsilon\to 1$ leads to a nontrivial fixed point in $3d$, applicable (for example) to critical phenomena in spin systems.\footnote{Alternatively, conformal bootstrap techniques were used in \cite{bootstrap3D} to study three-dimensional fixed points.}  Here, instead, we will send $\epsilon\to 0$ but introduce a mildly space- or time-dependent coupling (\ref{kappafour}) to obtain a perturbative critical point to very good approximation, with $\kappa\ll 1$ playing a role similar to $\epsilon\ll 1$ in the perturbative Wilson-Fisher fixed point.   

\subsection{The spacetime dependent case:\ critical phenomena in 4.0 dimensions}\label{subsec:scalar-timedep}

With translation invariance broken classically, quantum corrections will generate a richer $x$-dependent coupling $\tilde\lambda_0(x)$ than we started with, generalizing (\ref{minsub}).  This in particular involves terms with higher powers of $x$, which are more relevant than the term $\lambda (x \mu)^\kappa\phi^4$ that we began with.  It is also possible to start with a more general function of $x$ classically and cancel the $x$-dependent corrections, an analysis we will present in the following subsection.  For now we will simply work in the regime (\ref{xbound}) and extract the leading effect of $\kappa\simge 0$.  We will start by formulating the renormalization of our couplings.  The fixed point we find manifests itself in the scaling behavior of
correlation functions, as we will discuss in the example of QED below by studying the potential between charges.  
     
To start, let us consider the structure of the one-loop correction to the quartic coupling.  In position space, the one-loop diagrams  are of the form
\beq\label{oneloop}
\lambda^2\mu^{2(\epsilon+\kappa)}\int d^d{x}_+ \left(\int  d^d{ x}_-\frac{x^\kappa x'^\kappa}{({\vec x}_-^2+t_-^2)^{d-2}} \right)
\eeq
where $x_\pm=(x \pm x')$.
We are interested in the contributions to this integral which will produce poles in $\epsilon$ and hence contribute to the $\beta$ functional for $\lambda(x)$.  To extract that,
we can restrict our attention to the regime of the integral with $x_-\to 0$, in particular $|x_-|\ll |x_+|$.  This simplifies it to
\beq\label{oneloopsimple}
 \lambda^2\mu^{2(\epsilon+\kappa)}\int d^d{ x}_+ x_+^{2\kappa}\left(\int  d^d{ x}_-\frac{1}{({\vec x}_-^2+t_-^2)^{d-2}} \right)\,.
\eeq
Given this, computation of the poles reduces to that in the standard spacetime-independent $\phi^4$ theory and we can immediately write down the generalization of (\ref{minsub}):
 \beq\label{minsubkappa}
\tilde\lambda_0(x) =\mu^\epsilon \left[\lambda \mu^\kappa x^\kappa+\frac{3\lambda^2\mu^{2\kappa}x^{2\kappa}}{16\pi^2}\frac{1}{\epsilon}+\sum_{n=3}^\infty\lambda^n \mu^{n\kappa} x^{n\kappa} \sum_{i=1}^{n-1} \frac{c_{ni}}{\epsilon^i}\right]\,.
\eeq   
            
We can again use the fact that the bare coupling $\tilde\lambda_0(x)$ does not depend explicitly on $\mu$ (but only on the cutoff $\epsilon$ and the classical coupling) 
to derive the $\beta$ function(al) as before, generalizing (\ref{betaWF}).  At the level we are working so far, it will not vanish identically for any value of the constant parameter $\lambda$ because of the more general $x$ dependence we have generated, as expected.  
We will discuss this $x$-dependence further below, but for now let us extract the leading nontrivial contribution we will need in the regime (\ref{xbound}).  

Anticipating at least an approximate fixed point with $\lambda_*$ of order $\kappa$, we can write (\ref{minsubkappa}) as
 \beq\label{minsubkappasmall}
\tilde\lambda_0 =\mu^{\epsilon+\kappa}x^\kappa \lambda +\frac{3\lambda^2\mu^\epsilon}{16\pi^2}\frac{1}{\epsilon}+\dots
\eeq
 This gives
\beq\label{betakappa}
\beta = \mu\frac{\partial\lambda}{\partial\mu}=-(\epsilon+\kappa)\lambda  + \frac{3\lambda^2}{16\pi^2} +\dots,
\eeq
where $\dots$ includes terms of order $\lambda^3\sim\kappa^3$ and higher, as well as small $x$-dependent corrections from expanding $x^\kappa$.  This implies at least an approximate fixed point
\beq\label{lambdastar}
\lambda_* \approx \frac{16\pi^2}{3}\kappa
\eeq
as we set $\epsilon=0$.  It is straightforward to check explicitly up through order $\lambda^3\sim\kappa^3$ using (\ref{minsubkappa}) and standard results (see e.g.\ \cite{Kleinert}) that the pole terms cancel.  This includes a set of terms of order $\kappa\lambda^2/\epsilon$ that were not present in the ordinary spacetime-independent theory, but which still cancel.  This is consistent with the fact that the ultraviolet divergences in our theory are the same as in the usual theory, so precisely the same set of counterterms are required.    

This is related to the fact -- which is important to stress -- that this cancellation in the $\beta$ function occurs with respect to a transformation where the relative and total distances are scaled in the same way, namely $x_{\pm} \to x_{\pm}/b$.  Had we chosen to scale only $x_-$ while keeping $x_+$ fixed, the spacetime dependence $x^\kappa$ would not have shifted the dimension of the quartic coupling, and so there is no fixed point under that transformation. We will return to this point in \S \ref{subsec:force}, where we will show this effect in a more physical way.

So far this gives a fixed point valid over a finite but exponentially large range of scales $e^{-1/\kappa}\ll x\mu\ll e^{1/\kappa}$ (\ref{xbound}).
 It may be worth emphasizing that the coordinate and parameter choices we have made in expressing this relation are natural, since one could trivially redefine variables to produce an exponential relation from a power law or vice versa.  The coordinate $x$ measures the proper distance in spacetime, as opposed to say $w=\,\log(x\mu)$ (which satisfies only a power law relation $|w|\ll 1/\kappa$).  The parameter $\kappa$ here appears linearly in the $\beta$ function and the fixed point coupling, much like the inverse of the number of flavors $N_f$ in a system controlled by a $1/N_f$ expansion.  So it is fair to say that (\ref{xbound}) covers an exponentially large range of scales in terms of the natural parameters of the theory.  

Nonetheless, it is interesting to ask whether the $x$-dependence could be cancelled by a more general choice for our classical spacetime-dependent coupling.  This problem turns out to have a simple solution derived from the RG structure of the theory, as we will discuss next by including a systematic expansion in powers of $\kappa \,\log(x\mu)$. 

\subsubsection{General $x$-dependence and exact fixed points}
\label{subsubsec:exact}

To systematically analyze the $x$-dependence in our coupling, let us rewrite (\ref{minsubkappa}) as
\beq\label{minsubkappa2}
\tilde\lambda_0(x) =\mu^\epsilon \left[\lambda(x)+\frac{3\lambda(x)^2}{16\pi^2}\frac{1}{\epsilon}+\sum_{n=3}^\infty \lambda(x)^n \sum_{i=1}^{n-1} \frac{c_{ni}}{\epsilon^i}\right]
\eeq        
and expand
\beq\label{logexp}
\lambda(x)=\lambda_0 + \lambda_1\kappa \,\log(x\mu)+\lambda_2(\kappa \,\log(x\mu))^2+\dots=\sum_{m=0}^\infty \lambda_m (\kappa \,\log(x\mu))^m
\eeq
This generalizes the coupling we considered above, where we had $\lambda_0=\lambda_1=\lambda$ to the leading nontrivial order required to exhibit the cancellation in the $\beta$ function described above.
Now to generalize that, we would like to derive the renormalization group $\beta$ functions for all of the dimensionless couplings (in our basis (\ref{logexp}), the $\lambda_m$'s).     
We can again use the fact that $\partial\tilde\lambda_0(x)/\partial\mu = 0$ to find
\bea\label{bftn}\nonumber
\frac{\mu\partial\lambda(x)}{\partial\mu} &=& \sum_{m=0}^\infty \left[\mu\frac{\partial\lambda_m}{\partial\mu}(\kappa \,\log(x\mu))^m+\kappa m\lambda_m(\kappa \,\log(x\mu))^{m-1}\right] \\
&=& \frac{-\epsilon\lambda(x)-\frac{3}{16\pi^2}\lambda(x)^2-\sum_{n=3}^\infty\lambda(x)^n\sum_{i=1}^{n-1}\frac{c_{ni}}{\epsilon^{i-1}}}{1+\frac{6\lambda(x)}{16\pi^2\epsilon}+\sum_{n=3}^\infty n\lambda(x)^{n-1}\sum_{i=1}^{n-1}\frac{c_{ni}}{\epsilon^i}}\,.
\eea
This equation contains the $\beta$ function for each $\lambda_m$ in (\ref{logexp}).  At order $(\kappa \,\log(x\mu))^0$ we have
\beq\label{zero}
\mu\frac{\partial\lambda_0}{\partial\mu}=-\epsilon\lambda_0-\kappa\lambda_1 +\frac{3\lambda_0^2}{16\pi^2}+{\cal O}(\lambda_0^3)\,.
\eeq
Setting this to zero (and sending $\epsilon\to 0$), we obtain
\beq\label{lamzero}
\kappa\lambda_{1*}=\frac{3\lambda_{0*}^2}{16\pi^2}+{\cal O}(\lambda_0^3)
\eeq
This reproduces our earlier result (\ref{lambdastar}), noting that there we had $\lambda_0=\lambda_1$.  

Now let us consider the terms that depend nontrivially on $\kappa \log(x\mu)$.  At order $(\kappa \log(x\mu))^1$ we have
\beq\label{lamone}
\mu\frac{\partial\lambda_1}{\partial\mu}=-\epsilon\lambda_1 -2\kappa\lambda_2 +6\frac{\lambda_0\lambda_1}{16\pi^2}+\sum C_{\dots}\lambda_0\dots\lambda_0\lambda_1
\eeq
where we have been schematic in the last term, which expresses the fact that all contributions at this order come from products of $\lambda_m$'s whose indices add to 1.
Setting this to zero determines $\lambda_{2*}$ in terms of $\lambda_{1*}$ and $\lambda_{0*}$.  This pattern continues: at order $(\kappa \,\log(x\mu))^m$ we find
\beq
\mu\frac{\partial\lambda_m}{\partial\mu}=-\epsilon\lambda_m -(m+1)\kappa\lambda_{m+1} +\frac{3}{16\pi^2}\sum_{i=0}^m \lambda_i\lambda_{m-i}+\sum_{k\geq 3} C_{\dots}\lambda_{i_1}\lambda_{i_2}\dots\lambda_{i_k}
\eeq
where the last sum is restricted to $i_1+i_2+\dots+i_k=m$, so at the fixed point we determine $\lambda_{m+1*}$ in terms of the $\lambda_{n*}$ with $n\leq m$.  For the case $\lambda_{0*}=\lambda_{1*}$ (which we can always achieve by a redefinition of $\kappa$), one can show that all $\lambda_{m*}$ are equal to $16\pi^2\kappa/3$ up to higher order corrections in $\kappa$.  Thus the expansion \eqref{logexp} can be resummed to
\beq\label{lamxfix}
\lambda_*(x)=\frac{16\pi^2}{3}\frac{\kappa}{1-\kappa \log(x\mu)} \,.
\eeq
One can verify this by noting that it satisfies the RG equation \eqref{bftn} at leading order in $\lambda(x)$ with all $\lambda_m$'s being constant.  This is valid as long as $\lambda_*(x)$ stays within the perturbative regime, which is $x\mu\ll e^{1/\kappa}$ from \eqref{lamxfix}.  

This solution (\ref{lamxfix}) is precisely the running coupling of the translationally invariant theory, with scale replaced by $1/x$.  That can be seen from the fact that the condition for a fixed point is that the $\lambda_m$ coefficients in (\ref{logexp}) be independent of $\mu$.  Given that, $\mu$ only appears in the combination $x\mu$, and the RG equation \eqref{bftn}\ gives the solution just noted with $1/x$ playing the role of the scale.  This holds to all orders, and is limited only by the breakdown of perturbation theory when this spatially running coupling grows sufficiently large.      
It is interesting to consider the generalization of this to other theories with flows in which the coupling remains bounded;
one can for example apply this method to large-N theories to gain control over such trajectories.

Next, let us analyze the stability of this fixed point under variations the coupling, deforming $\lambda(x)\to \lambda_*(x)+\delta\lambda(x)$.  The leading $\beta$ function for this deformation is given by
\beq\label{betadelta}
\mu\frac{\partial\delta\lambda}{\partial\mu}{\Big |}_{\text{fixed} ~ x} = \frac{3}{8\pi^2}\lambda_*(x)\delta\lambda(x)
\eeq
This is solved by linear combinations of terms of the form
\beq\label{pertsol}
\frac{1}{(1-\kappa \,\log(x\mu))^2}\left(\frac{\mu}{M}\right)^r e^{-r \,\log(x\mu)}
\eeq
indexed by $r$, where $M$ is an integration constant rendering the equation dimensionally consistent.  It is natural to consider only contributions with $r\le {\cal O}(\kappa)$ since the translation
symmetry is only broken by contributions of order $\kappa \log(x\mu)$.  
Expanding this in powers of $\log(x\mu)$ determines the flow in the couplings $\delta\lambda_m$ (\ref{logexp}).  We see from this that there are relevant and irrelevant perturbations, depending on the 
sign of $r$.    

We should emphasize again that just at the level of the analysis above in \S\ref{subsec:scalar-timedep}\ we obtained critical behavior to very good approximation in the exponential window (\ref{xbound}) without needing to tune away all of the relevant perturbations in $\lambda(x)$.  There we needed to tune one number $\lambda\to\lambda_*$ to achieve this.         

Of course, in this scalar field theory an exact fixed point also requires a functional tune of the mass squared term (though one still obtains a very nearly critical theory over the range (\ref{xbound}) by tuning a single mass parameter).  That tuning is avoided in gauge theories with fermionic matter enjoying a global chiral symmetry to which we turn in the next section.

\section{$\kappa$QED and $\kappa$QCD}\label{sec:kappaQED}

In \S \ref{sec:4dscalar} we exhibited a new type of perturbative fixed point in the four-dimensional $\phi^4$ theory, which ensues after making the coupling slowly spacetime dependent. In fact, as we argued in the introduction, such perturbative critical points are more general:\ starting from a marginal interaction we can deform the coupling with a mild spacetime dependence to obtain a fixed point. In this and the following sections we will study other applications of this idea, including theories with supersymmetry or gauge interactions, as well as models in lower dimensions.

In this section we obtain an example from four dimensional QED with massless charged matter.  Without spacetime dependence, that theory is infrared free as a result of the one-loop screening effect of the charged matter.  By introducing  a  weakly spacetime-dependent coupling, we obtain a fixed point of the same sort we described above in $\phi^4$ theory, which we will call $\kappa$QED.\footnote{This paper is made possible by readers like you.}  Since the idea is the same, we will be briefer in this section. 

Consider the gauge invariant action
\beq\label{kQED}
S=\int d^4 x\left\{ -\frac{1}{4 e^2} F^2 + i\bar\psi\gamma^\mu{D_\mu}\psi \right\}
\eeq
where again we can consider the special case $e(x)=e x^\kappa\mu^\kappa$ in the regime (\ref{xbound}), or expand more generally as in  (\ref{logexp}), writing
\beq\label{alphalogexp}
e(x)=e_0+e_1 \kappa \,\log(x\mu)+e_2(\kappa \,\log(x\mu))^2+\dots.
\eeq
Similarly to the previous case (\ref{betakappa}), at leading order in the expansions in $\lambda$ and $\kappa \,\log(x\mu)$ we find 
\beq\label{betakappae}
\beta=-\kappa e + \frac{e^3}{12\pi^2}+\dots
\eeq   
where the second term is the usual one-loop beta function for massless QED.  Therefore we find a fixed point with, to good approximation,
\beq\label{estar}
e_*^2=4\pi\alpha_*=12\pi^2\kappa \,.
\eeq
As in the case of $\lambda\phi^4$ theory, the exact fixed point coupling $e^2(x)$ is given by the running coupling of the original theory with scale replaced by $1/x$.  

Similarly for massless QCD \cite{QCD}, we may choose $\kappa<0$ and obtain a UV fixed point $g_*^2\sim -\kappa$.  In order to stay in the perturbative regime in this case, we require $\Delta x$ to be shorter than the QCD scale:\ the range \eqref{Landau} becomes
\beq\label{LamQCD}
\Delta x\mu \ll e^{c/g^2} \,.
\eeq
The bound \eqref{xbound} on $x$ still applies with $\kappa$ replaced by $|\kappa|$.

\subsection{Force between charges}\label{subsec:force}

Having worked out the RG structure of $\kappa$QED, let us next consider the structure of physical observables.  Consider for example a pair of charges at points $x_1=(x_++x_-)/2$ and $x_2=(x_+-x_-)/2$.  In ordinary massless QED, the running of $\alpha$ leads to a potential energy between charges of the form
\beq\label{QEDpot}
V(x_-)\sim \frac{\alpha}{x_-}\left(1-\frac{\alpha}{3\pi} \log(x_-^2\mu^2)\right)+\dots
\eeq
In particular, if we rescale the coordinates by $x\to x/b$, the classical (Coulomb) potential scales like $V_\text{Coulomb}\to  bV_\text{Coulomb}$, but at the quantum level this simple scaling is violated by the logarithmic term.
In our case there is additional $x_+$ dependence
\beq\label{kQEDpot}
V(x_-)\sim \frac{\alpha}{x_-}\left(1+2\kappa \,\log(x_+\mu)-\frac{\alpha}{3\pi} \,\log(x_-^2\mu^2)\right)+\dots
\eeq
At our fixed point, we have $\alpha=\alpha_*=3\pi\kappa$ to good approximation.  For this value of the coupling, the potential (\ref{kQEDpot}) scales like $V\to bV$ if we rescale all coordinates.  

This exhibits the effect of the cancelling $\beta$ function in our theory on a basic physical observable.  It also illustrates the distinction between this scaling $x_\pm\to x_\pm/b$ and a scaling of only $x_-\to x_-/b$, with  $x_+$ fixed.  In a translationally invariant theory, the distinction does not come up, but here it is important.  Clearly if we scale only $x_-$ without scaling $x_+$, the potential exhibits a log dependence on this scale and hits a Landau pole at short distance $x_-$.     
This is related to the fact that although $\alpha$ behaves like the Wilson-Fisher $\epsilon$ in its effects on our $\beta$ function, it does not play the same role in cutting off the theory and rendering it super-renormalizable.

\subsection{Digression:\ matching at the electron mass scale}\label{subsec:scales}

For amusement, let us now explore the possibility of this fixed point arising in the history of the observable Universe.
For massive $\kappa$QED, we lose the second term in (\ref{betakappae}) at distance scales longer than the inverse electron mass.  
Given the relation (\ref{xbound}), let us check the scales involved given the measured value $\alpha\approx 1/137$.  That is, let us suppose (just for fun) that the fixed point we have found pertains in the real world, at short distances compared to the electron Compton wavelength $1/m_e$.  Matching at that scale, we then have at longer distances
\beq\label{ereal}
\alpha=3\pi\kappa (m_e x)^{2\kappa}
\eeq 
or equivalently
\beq\label{alphareal}
\alpha(x)\approx \frac{1}{137}(m_e x)^{2/(3\pi\times 137)}
\eeq
Let us estimate the spacetime variation of $\alpha$ that this represents.  In particular, let us check how the factor $(m_e x)^{2/(3\pi\times 137)}$ behaves
at the the longest spacetime scales in the observable universe, namely $x\sim 10^{60}/M_P$ where $M_P\sim 10^{19}$ GeV is the Planck mass scale.   
This is
\beq\label{xfactor}
\left(\frac{10^{60}}{10^{19}\textrm{GeV}}\times 0.5\times 10^{-3}\textrm{GeV}\right)^{2/(3\pi\times 137)}\sim 1.1
\eeq
At face value this would represent a $10\%$ variation of $\alpha$, much bigger than the current bounds and claimed detections.  
It would be interesting to check for applications of our fixed points in lab systems with more tunable parameters; we will comment on this for statistical systems in three dimensions next.

\section{The three-dimensional case and statistical physics}\label{sec:3d}

Having established the existence of the new class of perturbative fixed points (\ref{lambdastar}) in four-dimensional theories, we will next apply the same technique to derive three-dimensional examples.  One motivation is that this is the dimensionality relevant for potential realizations in laboratory statistical mechanical systems. The fixed point found in \S \ref{sec:4dscalar} (and those that will be analyzed in \S\ref{sec:susy}) have close analogues in three dimensions. Here we will focus on the $\phi^6$ theory, which arises in the low energy effective field theory limit of some condensed matter systems.  

In more detail, consider a Euclidean $O(N)$ model for $N$ real scalars $\phi_i$ with bare Lagrangian
\be
\mathcal{L} = \frac{1}{2} (\partial \phi)^2 +\frac{1}{2} m_0^2 \phi^2 + \frac{\lambda_0}{4!} (\phi^2)^2+ \frac{g_0}{6!} (\phi^2)^3\,.
\ee
This theory describes various systems whose phase diagrams feature tricritical points, such as polymers, metamagnets and $^{3}\text{He}-{^{4}\text{He}}$ mixtures, etc.; see e.g.~\cite{hager, griffits}. The bare parameters $m_0$ and $\lambda_0$ will be tuned to exactly cancel the renormalized mass and quartic interactions. In the $^{3}\text{He}-{^{4}\text{He}}$ case, for instance, this corresponds to tuning the temperature and relative chemical potential. The renormalized action in $d=3-\epsilon$ dimensions is then
\be
S= \int d^dx\,\left[\frac{1}{2} Z_\phi (\partial \phi_R)^2 +\mu^{2\epsilon} Z_g\,\frac{g}{6!} (\phi_R^2)^3\right]\,,
\ee
where $g_0 =\mu^{2\epsilon} \frac{Z_g}{Z_\phi^3} g$. Using the results of~\cite{hager} for $Z_g$ and $Z_\phi$ and requiring that $g_0$ be independent of the RG scale $\mu$ gives the beta function
\be\label{eq:beta-phi6}
\mu \frac{\partial g}{\partial \mu}= -2 \epsilon g + \frac{22+3N}{240 \pi^2} g^2 + \mc O(g^3)\,.
\ee
Here the lowest two-loop correction proportional to $g^2$ comes from $Z_g$; higher order contributions may be found in~\cite{hager2}. This gives a three-dimensional analogue of the Wilson-Fisher fixed point, with $g$ becoming marginally irrelevant in the limit $\epsilon \to 0$.

Now we add a mild power-law dependence, starting from
\be
g \to (x \mu)^\kappa g\,
\ee
with $0 < \kappa \ll 1$, which shifts (\ref{eq:beta-phi6}) by $-\kappa g$. Taking the limit $\epsilon \to 0$, we find a perturbative fixed point at
\be
g_* \approx \frac{240 \pi^2}{22+3N} \kappa\,.
\ee
As above, we obtain the complete fixed point coupling function $g(x)$ as in \S\ref{subsubsec:exact}.  
 It is straightforward to extract the critical exponents predicted by our theory.  

It would be interesting to understand how to engineer the $x$-dependence at a microscopic level in real systems.  Presumably if one could introduce translation breaking in a lab system (perhaps with spacetime variation in pressure or another accessible parameter), this would propagate down to the effective theory and lead to couplings of our form. 
Another interesting question is whether some materials come naturally with gradients which could affect their scaling behavior and critical exponents as in our theory. 

\section{Nonlinear $\sigma$-models in two dimensions:\ fixed points for Ricci-curved target spaces}\label{sec:sigma-model}

Moving to lower dimension, let us remark on another interesting class of examples of this phenomenon.  Consider $\sigma$-models in two-dimensional QFT,
\beq\label{sigmaaction}
S =\frac{1}{4\pi}\int d^2x G_{ij}(Y, x)\partial_\mu Y^i\partial^\mu Y^j
\eeq
where $Y^i$ denote the embedding coordinates into the target space of the sigma model.  It is a standard result \cite{Friedan}\ that the one-loop beta function is proportional to the Ricci tensor, and that  the theory is asymptotically free for a positively curved target space and infrared free for a negatively curved target space.  

Including a spacetime dependent coupling $G_{ij}(Y,x)\approx (x \mu)^\kappa \hat G_{ij}(Y)$, the beta functional for this theory takes the form 
\beq\label{sigmabeta}
\beta^{\hat G}_{ij}=-\kappa {\hat G}_{ij}-{\hat R}_{ij}
\eeq
up to higher order corrections.  Here ${\hat R}_{ij}$ is the Ricci tensor for the target space manifold (in a convention where it is positive for a sphere).  
As in the previous examples, this enables us to find new perturbative fixed points under perturbative control
for an exponentially large range of scales.  From (\ref{sigmabeta}) we see that these are given by Einstein spaces:
\beq\label{Gstar}
{\hat R}_{ij}({\hat G}_*)=-\kappa {\hat G}_{ij*}
\eeq  
For negatively curved target spaces,\footnote{which are by far the most generic among manifolds} this produces a new class of fixed points to very good approximation starting from a theory which would normally flow toward weak coupling in the infrared.  The flow will be in the opposite direction for the positively curved spaces.  All of this is in good analogy with the discussion above for QED and QCD.  It would be interesting to explore potential applications to statistical systems in two spatial dimensions or to string theory in curved spaces.    

\section{Additional examples from conformal perturbation theory and supersymmetry}\label{sec:susy}

In this section, we will generalize our mechanism in a way that will apply to certain strongly coupled theories.  Then we will develop some specific examples using the theoretical tools of supersymmetry and holographic duality.  

\subsection{General framework}\label{subsec:general}

We can generalize our mechanism as follows. Consider a CFT with a marginal operator $\mc O$; upon deforming
\be\label{eq:Odef}
S_{CFT} \to S_{CFT} + \int d^d x\,\lambda\,\mc O\,,
\ee
$\mc O$ generically becomes either relevant or irrelevant;\footnote{We do not consider the special case of exactly marginal operators.} such operators are known as marginally relevant or irrelevant respectively. The CFT need not be weakly coupled, but we require $\lambda \ll 1$ so that conformal perturbation theory can be applied.  This leads to (schematically)
\be
\mu \frac{\partial \lambda}{\partial \mu} = c\,\lambda^2 + \ldots\,,
\ee
where $c$ is proportional to the $\mc O$ 3-point function.  

Starting from such a CFT, we will now deform it by $\mc O$ but with a spacetime dependent coupling
\be\label{eq:perturb-st}
S= S_{CFT}+ \int d^dx\,\lambda (x\mu)^\kappa\,\mc O
\ee
with $|\kappa| \ll 1$. The main consequence of this is that $\lambda$ acquires a small dimension $\kappa$. Over a parametrically large range of scales $e^{-1/|\kappa|} \ll x\mu\ll e^{1/|\kappa|}$,
with also a lower/upper bound on coordinate differences applicable in the marginally irrelevant/relevant case as in (\ref{Landau}) or (\ref{LamQCD}), 
the beta function is then well approximated by
\be\label{eq:beta-st}
\mu \frac{\partial \lambda}{\partial \mu} = -\kappa \,\lambda+ c\,\lambda^2 + \ldots\,.
\ee
This yields a perturbative fixed point at
\be\label{eq:lambda-fixed}
\lambda_* \approx \frac{\kappa}{c}\,.
\ee
As in the previous examples, we obtain an exact fixed point $\lambda_*(x)$ as in \S\ref{subsubsec:exact}.  

This construction is very general:\ all that is needed is a CFT with a marginal operator, and the new fixed point is obtained by balancing the mild spacetime dependence (\ref{eq:perturb-st}) against perturbative corrections. The existence of these fixed points can be consistently established using conformal perturbation theory and does not require the full theory to be weakly coupled.

In the remainder of this section we will discuss supersymmetric theories which, besides providing a different class of examples, will extend the results of \S \ref{sec:4dscalar} in two interesting directions. First, reaching the fixed point will no longer require tuning the mass term to cancel quantum corrections, because these are absent to leading order. And secondly, with the help of supersymmetry and conformal perturbation theory, we will extend the construction into the strongly coupled regime. These aspects will be illustrated, respectively, with the Wess-Zumino model and with superconformal field theories deformed by marginally irrelevant operators.\footnote{Of course, such fixed points can also exist in nonsupersymmetric CFTs, but supersymmetry is helpful in identifying the appropriate deformations.}

\subsection{Static Wess-Zumino model}\label{subsec:staticWZ}

Let us begin by showing that the Wess-Zumino model admits a fixed point similar to that of the $\phi^4$ theory (and Yukawa theory), without the need to fine-tune the mass. We first review the Wess-Zumino model at one loop, and then move to the spacetime dependent case.

In four dimensions, the theory contains a complex scalar $\phi$ and a Weyl fermion $\psi$ that can be grouped into a chiral superfield $\Phi$. The interactions are encoded into a superpotential
\be
W= \frac{1}{6}\, \hat y \,\Phi^3\,,
\ee
which gives rise to Yukawa and quartic interactions $L \supset - \frac{1}{2}\hat y \phi \psi \psi - c.c. - \frac{1}{4}|\hat y|^2 |\phi|^4$.

In this theory, the superpotential is not renormalized and at quadratic order the only renormalization comes from the K\"ahler potential,
\be
K = Z(\mu) \Phi^\dag \Phi\,.
\ee
Changing to canonically normalized fields, the physical coupling $y$ is related to the holomorphic $\hat y$ by $y = Z(\mu)^{-3/2} \hat y$. 
Since $\hat y$ is not renormalized, the beta function for $y$ becomes
\be\label{eq:WZbetay}
\beta_y = \frac{3}{2}\gamma y\,,
\ee
with the anomalous dimension defined as $\gamma \equiv - \frac{\partial \log Z(\mu)}{\partial \log \mu}$.\footnote{We remind the reader that the dimension of $\Phi$ is given in terms of $\gamma$ by $\Delta(\Phi) = \frac{d-2}{2}+\frac{\gamma}{2}$.} In perturbation theory, the anomalous dimension at one loop is as in Yukawa theory, giving (see e.g.\ \cite{Martin})
\be\label{eq:gammaWZ}
\gamma = \frac{|y|^2}{16 \pi^2} + \mc O(|y|^4)\,.
\ee
(Recall that the unitarity bound requires $\gamma>0$).

We are now ready to discuss the possible fixed points of this theory. In $d$ dimensions, the scaling dimension of $\hat y$ is $(4-d)/2$, so the dimensionless coupling
\be
\eta \equiv \mu^{\frac{d-4}{2}} y 
\ee
has beta function
\be
\beta_\eta = \frac{1}{2}\eta \left(-(4-d) + 3 \gamma \right)\,.
\ee
Therefore, in $d=4-\epsilon$ dimensions the theory has a perturbative fixed point
\be
\gamma_* = \frac{\epsilon}{3}\;\Rightarrow\; |y_*|^2= \frac{16 \pi^2}{3} \epsilon\,.
\ee
As with the Wilson-Fisher fixed point, one can study this theory in the limit $d \to 3$ using the $\epsilon$ expansion.

\subsection{Perturbative fixed point from spacetime dependence}\label{subsec:WZ-timedep}

When $d \to 4$, the coupling becomes marginally irrelevant and the previous fixed point is lost. We will now show that adding spacetime dependence via
\beq\label{ykappa}
\hat y\to \hat y x^{\kappa/2},
\eeq
with $\kappa \ll 1$ leads to a perturbative fixed point. (We choose the power to be $\kappa/2$ so that the $x$-dependence in the $\phi^4$ potential agrees with that of \S \ref{sec:4dscalar}.)   One can consider this modification directly in components, or obtain a spacetime-dependent superpotential via a chiral superfield which varies with $x$, breaking supersymmetry spontaneously.  

Of course the breaking of translation invariance and hence supersymmetry by the spacetime dependent coupling implies that the superpotential is not exactly protected from quantum corrections. In particular, a mass term can be generated, and $\hat y$ will be renormalized.  What supersymmetry introduces is a natural cutoff on the mass in the UV.  In the infrared, the situation is similar to our previous perturbative examples, where we obtain a fixed point to very good approximation
in the regime (\ref{xbound}).  In that regime, the supersymmetry-breaking contributions to the scalar and fermion propagators are negligible.   

The coupling $\hat y$ has classical dimension $\kappa/2$, so we look for a fixed point of the dimensionless physical coupling
\be
\eta \equiv \frac{Z^{-3/2} \hat y}{\mu^{\kappa/2}}\,.
\ee
In the limit where supersymmetry-breaking effects can be ignored, $\beta_{\hat y}=0$ and
\be
\beta_\eta = \frac{1}{2}\,\eta (-\kappa + 3 \gamma)\,.
\ee
Since the unitarity bound sets $\gamma>0$, choosing $0<\kappa \ll 1$ obtains a perturbative fixed point 
\be
\gamma_* \approx \frac{\kappa}{3} \,.
\ee
The one-loop result (\ref{eq:gammaWZ}) is valid in this regime, so the value of the coupling becomes
\be\label{eq:etafixed}
|\eta_*|^2 \approx \frac{16 \pi^2}{3} \kappa
\ee
up to subleading $x$-dependent corrections. This is the same result that we found in (\ref{lambdastar}), but the physical origin is somewhat different. In the present case there are both bosonic and fermionic contributions, and the fixed point arises from balancing the classical dimension induced by spacetime dependence against the anomalous dimension. In contrast, in the nonsupersymmetric example the contribution from the anomalous dimension was subleading.

\subsection{Spacetime dependent fixed points from superconformal field theories}\label{subsec:scft}

So far we analyzed fixed points that are obtained by adding spacetime dependence to weakly coupled theories. However, as we discussed in \S \ref{subsec:general}, this approach is in principle more general:\ it can apply to strongly coupled CFTs as long as we can do conformal perturbation theory. A difficulty in making this concrete lies in identifying a marginal operator $\mc O$ and then determining whether it becomes marginally relevant or irrelevant after adding the deformation (\ref{eq:Odef}). This can be accomplished in a wide variety of superconformal field theories.

A basic result that we can use to identify marginally irrelevant couplings was derived in \cite{Green:2010da}\ using global symmetries.
To start, let us briefly review how the static result of~\cite{Green:2010da} comes about. Consider a 4d SCFT with a chiral operator $\mc O$ that is marginal and charged under a global symmetry with current $J$. For simplicity we assume a $U(1)$ symmetry and denote the charge of $\mc O$ by $q$; similar results hold in the nonabelian case as well. The OPE of this operator is of the form
\be\label{OOJ}
 \mc O(x) \mc O^\dag(0)  = \frac{1}{|x|^6}+ \frac{\gamma q}{|x|^4} J(0) + \ldots\,,
\ee
where $\langle J(x) J(0)\rangle = \gamma/|x|^4$. A superpotential deformation $W_{CFT} \to W_{CFT}+ \hat \lambda \mc O$
then produces a logarithmic correction to the action of the form $\int d^4 \theta \,\Delta K = \int d^4 \theta\, Z J$. At lowest order in conformal perturbation theory, this arises from bringing down two powers of the deformation and using (\ref{OOJ}) 
\be
\int d^4x\, d^2 \theta\, \hat \lambda \mc O(x) \,\int d^2 \bar \theta\, \hat \lambda^\dag \mc O^\dag (0)\,.
\ee
The beta function for $Z$ is then
\be\label{eq:betaZ}
\mu \frac{\partial Z}{\partial \mu}=-2\pi^2 \gamma q |\hat \lambda|^2+ \ldots
\ee
As explained in \cite{Green:2010da}, 
this change can be absorbed into a redefinition of the coupling $\lambda \equiv \hat \lambda - \frac{1}{2} \gamma q\,\lambda\,Z$ which, from (\ref{eq:betaZ}), is marginally irrelevant:
\be\label{eq:strong-beta}
\mu \frac{\partial \lambda}{\partial \mu} = \pi^2 \gamma q^2\,\lambda |\lambda|^2\,.
\ee

This is all that we need to understand what happens when we deform the theory by a spacetime-dependent interaction
\be
W= W_{CFT}+ \lambda (x\mu)^{\kappa/2} \mc O\,.
\ee
Again, one can consider this modification directly in components, using supersymmetry to control the original unperturbed SCFT; potentially one can obtain such a spacetime-dependent superpotential via a very massive background chiral superfield which varies as $x^\kappa$.
As before, for $0 < \kappa \ll 1$ this gives a small positive dimension to the coupling, which can be balanced against (\ref{eq:strong-beta}) to yield a fixed point
\be\label{eq:scft-fp}
|\lambda_*|^2 \approx \frac{1}{2\pi^2}\,\frac{\kappa}{\gamma q^2}\,.
\ee
Again, there is a simple solution for the full $x$-dependence of the fixed point coupling as in \S\ref{subsubsec:exact}.  

There are many examples of this general result. A weakly coupled case is given by the WZ model of \S \ref{subsec:staticWZ}. There the SCFT is just a free theory of a single chiral superfield $\Phi$, and the marginal operator is $\mc O=\Phi^3$. This field is charged under a $U(1)$ symmetry that rotates it, so after deforming the theory by $W= \lambda \mc O$, it becomes irrelevant. The fixed point of \S \ref{subsec:WZ-timedep} sourced by spacetime dependence is in agreement with (\ref{eq:scft-fp}).

Among the zoo of known SCFTs, let us mention one interesting strongly coupled example ~\cite{Klebanov:1999tb}\ with a rich global symmetry structure.  This theory has a superpotential
$$
W_{CFT} = h \epsilon^{ik} \epsilon^{jl} \,\tr(A_i B_j A_k B_l) 
$$
where $A$ and $B$ are bifundamentals of the $SU(N) \times SU(N)$ gauge group of the theory, and $i, j$ are global $SU(2)$ indices. The operator $\tr(A_i B_j A_k B_l)$ (with a specific choice of $i,j,k,l$) is marginal at the fixed point, and it is charged under the $SU(2) \times SU(2)$ global symmetry. So it becomes irrelevant after adding it as a deformation~\cite{Green:2010da}. Therefore, adding appropriate spacetime dependence
\be
W = h \epsilon^{ik} \epsilon^{jl} \,\tr(A_i B_j A_k B_l) + \lambda_*(x)\tr(A_i B_j A_k B_l)
\ee
leads to a new  fixed point (\ref{eq:scft-fp}). This example, being both supersymmetric and holographic, suggests that our mechanism may also have a GR realization in AdS solutions.  It would be interesting to consider the scalar field solution corresponding to this deformation and its relation to the scale-radius correspondence in gauge/gravity duality.

\section{Discussion}\label{sec:discussion}

In this paper, using mild breaking of translation symmetry, we have introduced novel theories which are scale invariant over an exponentially large range of scales (\ref{xbound}).  They can be thought of as physical analogues of Wilson-Fisher fixed points, and arise in standard perturbative field theory as well as in conformal perturbation theory.  Spacetime dependence in the couplings of the form $\lambda\to\lambda (x\mu)^\kappa$ introduces a new term $-\kappa\lambda$ into the $\beta$ function for $\lambda$ which can cancel higher order contributions that would otherwise render the coupling marginally relevant or irrelevant.  The exact spacetime-dependent fixed point coupling is obtained by choosing $\lambda_*(x) = \bar\lambda(1/x)$, where $\bar\lambda$ is the running coupling of the original translationally invariant theory.  

Many of the examples we discussed are based on simple field theories with minimal field content (no large flavor or color expansions, nor supersymmetry, being required when control is afforded by perturbation theory).  This raises the possibility of real-world applications, including potential generalizations of critical phenomena, a classic interface of quantum field theories and experiments.  It would also be interesting to compute transport coefficients in these theories.  In a different vein, in standard cosmology, the scale factor varied as a power of time during the matter- and radiation-dominated epochs, leading to shifts in the scaling dimension of the relevant terms in the action;\footnote{Marginal terms are not affected at leading order, FRW backgrounds being conformally flat.} the powers $\kappa=2/3,1/2$ in this case are not small, but the flow may be interesting to consider in cosmological calculations.  Slow-roll inflation can also mildly break time translation invariance, and may affect RG running for fields coupled to the inflaton.\footnote{This is a somewhat different effect from those captured in \cite{EFTinflation}, which gave an interesting RG framework for inflationary perturbations.}      
More formally, these results may aid in analyzing spacetime dependent backgrounds in string theory, including those built from AdS/CFT \cite{FRW}, the subject which led us to these considerations.  In general, most systems (including the universe as a whole) are not spacetime translation invariant, and we hope these results help in developing a systematic understanding of the consequences of this.

\section*{Acknowledgements}

We thank S.\ Kachru,  A.\ Perko, J.\ Polchinski and S.\ Shenker for helpful comments.  This research program was made possible in part by the National Science Foundation under grant PHY-0756174, by the Department of Energy under contract DE-AC03-76SF00515, and by readers like you. 


\bibliographystyle{JHEP}
\renewcommand{\refname}{Bibliography}
\addcontentsline{toc}{section}{Bibliography}
\providecommand{\href}[2]{#2}\begingroup\raggedright

\end{document}